\documentclass[preprint,12pt]{elsarticle}

\usepackage{lineno}

\usepackage[utf8]{inputenc}
\usepackage[english]{babel}
\usepackage{amssymb}
\usepackage{amsfonts}
\usepackage{amsmath}
\usepackage{txfonts}
\usepackage{graphicx}
\usepackage{ae,aecompl}
\usepackage{fancyhdr}
\usepackage{multicol}
\usepackage{layout}
\usepackage{graphicx}
\usepackage{multirow}
\usepackage{times}
\usepackage{textcomp}
\usepackage{cprotect}
\usepackage[dvipsnames]{xcolor}
\usepackage{enumitem}
\usepackage{xcolor}

\usepackage[outdir=./figures/]{epstopdf}
\usepackage[export]{adjustbox}

\usepackage{comment}

\usepackage{hyperref}
\hypersetup{%
    ,urlcolor=blue
    ,citecolor=blue
    ,linkcolor=blue
    }
    
\graphicspath{{./figures/}}

\biboptions{sort&compress}

\newif\ifAMStwofonts
\AMStwofontstrue

\begin{document}

\begin{frontmatter}

\title{Kinetic coupled tachyon: A dynamical system analysis}

\author[1,11,111]{Francesco Pace}
\ead{francesco.pace@unito.it}

\affiliation[1]{organization={Dipartimento di Fisica, Universit\`a degli Studi di Torino},
            addressline={Via P. Giuria 1}, 
            postcode={I-10125}, 
            city={Torino},
            country={Italy}}

\affiliation[11]{organization={INFN-Sezione di Torino},
            addressline={Via P. Giuria 1}, 
            postcode={I-10125}, 
            city={Torino},
            country={Italy}}

\affiliation[111]{organization={INAF-Istituto Nazionale di Astrofisica, Osservatorio Astrofisico di Torino},
            addressline={strada Osservatorio 20}, 
            postcode={10025}, 
            city={Pino Torinese},
            country={Italy}}

\author[2]{Alberto Rozas-Fern\'andez}
\ead{a.rozas@oal.ul.pt}

\affiliation[2]{organization={Instituto de Astrof\'{\i}sica e Ci\^{e}ncias do Espa\c{c}o, Universidade de Lisboa, OAL},
            addressline={Tapada da Ajuda},
            postcode={PT1349-018}, 
            city={Lisboa},
            country={Portugal}}

\author[3]{\"{O}zgen Tun\c{c} T\"{u}rker\corref{cor1}}
\ead{ozgen.turker@edu.ufes.br}

\affiliation[3]{organization={PPGCosmo, Universidade Federal do Espírito Santo},
            addressline={Av. Fernando Ferrari, 514},
            postcode={29075-910}, 
            city={Vit\'oria},
            state={ES},
            country={Brazil}}

\cortext[cor1]{Corresponding author}

\begin{abstract}
We present and examine a kinetically coupled tachyon dark energy model, where a tachyon scalar field $\phi$ interacts with the matter sector. More specifically, we deduce this cosmological setting from a generalised interacting dark energy model that allows for the kinetic term of the scalar field to couple to the matter species \textit{a priori} in the action. A thorough dynamical system analysis and its cosmological implications unveil the appearance of a scaling solution which is also an attractor of the system, thanks to a novel critical point, with a period of accelerated expansion thereafter. This new solution, not present in the uncoupled case, has the enticing consequence of alleviating the coincidence problem.  
\end{abstract}


\end{frontmatter}

\section{Introduction}
More than twenty five years have past since the discovery of the late-time acceleration of the Universe. However, the origin and nature of the energy source responsible for it, known as dark energy (DE), remains a mystery. On top of this, a dark matter (DM) component is needed to account for the structure of the Universe and its evolution. The leading cosmological model that fits the observations \cite{Planck:2018vyg} best is called $\Lambda$CDM \cite{Carroll:2000fy, Padmanabhan:2002ji}. Here $\Lambda$ is the cosmological constant, the simplest form of DE, and CDM stands for cold DM, a non-baryonic component which could possibly be made of weakly interacting massive particles, axions but also of primordial black holes. Nevertheless, this standard cosmological model suffers from the well-known cosmic coincidence \cite{Fitch:1997cf} and fine tuning \cite{Weinberg:1988cp,Padilla:2015aaa} problems. This has forced theoreticians to look for alternatives, such as those based on scalar fields or modifications of gravity on large scales. One of the most popular scalar field scenarios is given by the tachyon field $\phi$, whose Lagrangian density is given by
\begin{equation}
\mathcal{L}_{\rm tach}=-V(\phi)\, \sqrt{1+g^{\mu\nu}\partial_{\mu}\phi\partial_{\nu}\phi}\,, 
\label{tach}
\end{equation}
where $V(\phi)$ is the tachyon potential.
Many studies have shown that the tachyon can act as a source of dark energy with
different potential forms  \cite{Bagla:2002yn, Padmanabhan:2002cp,Gibbons:2002md,Frolov:2002rr,Copeland:2004hq, Calcagni:2006ge,Shao:2007zv}. Additionally, dynamical studies on the tachyon scenario have been carried out. For instance, a barotropic perfect fluid with the tachyon field and several potential forms was explored in \cite{Copeland:2004hq}, in \cite{Fang:2010zze} for the case of the potential with a general form and in \cite{Aguirregabiria:2004xd} for the inverse square potential. Further studies with other types of potential can be found in \cite{Guo:2003zf, Quiros:2009mz}.

A natural extension of the idea of using a scalar field as a dynamical DE scenario would be to consider a non-minimal coupling to the DM sector since, a priori, there is no reason that forbids it. A tantalising consequence, when performing a dynamical system analysis of these models, is the possibility of obtaining scaling solutions, which may alleviate the cosmic coincidence problem. In particular, models with a variety of couplings between the tachyon field and DM have been explored in the literature \cite{Herrera:2004dh, Gumjudpai:2005ry, Farajollahi:2011jr, Farajollahi:2011ym, Landim:2015poa, Ahmad:2015sna, Shahalam:2017fqt, Teixeira:2019tfi}.

In this work, following the proposal in \cite{Barros:2019rdv}, we consider that the tachyon scalar field $\phi$, with Lagrangian density $P(\phi,X)$, kinetically couples to matter, where $X \equiv -\frac{1}{2}\partial^{\mu}\phi \partial_{\mu}\phi$ is the kinetic term. Accordingly, we shall assume that the term $X$ can couple directly to the matter fields at the level of the action. The interaction is mediated by a
general function $f(\phi,X)$ in the matter action. The cosmological field equations are derived and solved for a specific model. We find that the inclusion of the coupling allows for an early scaling solution followed by a period of accelerated expansion, when the function $f$ depends on the kinetic term only.

Assisted by dynamical systems techniques, the influence of the
coupling on the overall cosmological dynamics is examined. The scaling regime, useful to alleviate the cosmic coincidence problem, arises due to the emergence of a new critical point that only appears when the coupling is active. 

This work is structured as follows: in Sect.~\ref{model} the general equations of the theory are presented along with some particular examples. In Sect.~\ref{KCT}, the kinetic coupled tachyon cosmological model is introduced with a specific kinetic coupling function. The dynamical system analysis is performed in Sect.~\ref{DS}, where we obtain the equations of motion in terms of some suitable dynamical variables. In addition, we examine the nature of the critical points and their cosmological implications. Finally, we conclude in Sect.~\ref{Concl}.

\section{The model}
\label{model}

Let us consider a Universe filled with two components, namely, a scalar field $\phi$ interacting with one matter fluid. This Universe lives in a four-dimensional spacetime manifold $\mathcal{M}$ endowed with a metric $g_{\mu \nu}$.
For this kinetic coupled dark energy model, the total action minimally coupled to Einstein gravity is given by
\begin{equation}\label{actiongral}
 \mathcal{S} = \int \mathrm{d}^4 x \sqrt{-g} \left[\frac{M_{\rm Pl}^2}{2} R + P(\phi,X) + {f}(\phi,X)\tilde{\mathcal{L}}_{\rm m}(g_{\mu\nu},\psi) \right]\,,
\end{equation}
where $M_{\rm Pl} \equiv 1/ \sqrt{8 \pi G}$ is the reduced Planck mass and $R$ is the Ricci scalar, which depicts the geometrical sector of the Universe and is thus expressed in terms of $g_{\mu \nu}$. The term $\psi$ accounts for the matter field while $f(\phi,X)$, which depends on the field and on the  kinetic term $X \equiv -\frac{1}{2}g^{\mu\nu} \partial_{\mu}\phi \partial_{\nu}\phi$, encodes the information on the coupling between the field $\phi$ and the matter fluid. Finally, the Lagrangian densities of the scalar field and the matter fluid are $\mathcal{L}_{\phi} = P(\phi,X)$ and $\tilde{\mathcal{L}}_{\rm m}$, respectively.

The Einstein field equations are obtained by varying the action in Eq.~\eqref{actiongral} with respect to the metric
\begin{equation}
\label{EE}
M^{2}_{\rm Pl}G_{\mu\nu} = T^{(\phi)}_{\mu\nu} + {f}\, \tilde{T}^{(m)}_{\mu\nu} + {f}_{,X} \tilde{\mathcal{L}}_{\rm m} \partial_{\mu}\phi \partial_{\nu}\phi\,,
\end{equation}
with
\begin{equation}
T_{\mu \nu}^{\, (\phi)} \equiv -2\frac{\delta \mathcal{L}_{\phi}}{\delta g^{\mu\nu}} + \mathcal{L}_{\phi}g_{\mu\nu}\,, \  \ \ \tilde{T}^{\rm (m)}_{\mu \nu} \equiv -2\frac{\delta \tilde{\mathcal{L}}_{\rm m}}{\delta g^{\mu\nu}} + \tilde{\mathcal{L}}_{\rm m} g_{\mu\nu}\,,
\label{Tc}
\end{equation}
being the energy-momentum tensors of the scalar field and the matter fluid, respectively, and $G_{\mu \nu}$ is the Einstein tensor while $R_{\mu \nu}$ is the Ricci curvature tensor. Hereafter, we shall use the notation $P(\phi,X)\equiv P$ and ${f}(\phi,X)\equiv {f}$ in order to simplify the expressions.

When considering couplings that also depend on the kinetic term of the scalar field, a new interaction term arises -- the last one on the right-hand side of Eq.~\eqref{EE} -- in the Einstein equations. As will be shown, this affects the contracted Bianchi identities, having therefore an impact on the overall dynamics of the system.

It is possible to write the field equations for the current model in a more intuitive form, by defining the object
\begin{equation}\label{flm}
\mathcal{L}_{\rm m}(g_{\mu\nu},\psi,\phi,X) \equiv f(\phi,X) \tilde{\mathcal{L}}_{\rm m}(g_{\mu\nu},\psi)\,,
\end{equation}
which represents an effective matter Lagrangian that concentrates the effects of the coupling. Thus, we have the following relation for the energy-momentum tensors
\begin{align}
T^{\rm (m)}_{\mu\nu} = &\, -2\frac{\delta \mathcal{L}_{\rm m}}{\delta g^{\mu\nu}} + \mathcal{L}_{\rm m}g_{\mu\nu}\,, \nonumber \\
= &\, {f}\,\tilde{T}^{\rm (m)}_{\mu\nu} -2 \tilde{\mathcal{L}}_{\rm m}  \frac{\delta {f}}{\delta g^{\mu\nu}}\,, \nonumber\\
= &\, {f}\, \tilde{T}^{\rm (m)}_{\mu\nu} + {f}_{,X} \tilde{\mathcal{L}}_{\rm m} \partial_{\mu}\phi \partial_{\nu}\phi\,,
\end{align}
where ${f}_{,X} \equiv \frac{\partial f}{\partial X}$, and the field equations, Eq.~\eqref{EE},  may be recast as
\begin{equation}\label{EEs}
 M^{2}_{\rm Pl}G_{\mu\nu} = T^{(\phi)}_{\mu\nu} + T^{\rm (m)}_{\mu\nu}\,.
\end{equation}

Variation of the action in Eq.~\eqref{actiongral} with respect to $\phi$ yields the equations of motion for the scalar field (see Appendix A for further details)
\begin{equation}\label{final}
P_{,\phi} + P_{,X} \square\phi - 2 P_{,X\phi}X - P_{,XX}\nabla^{\mu}\phi\nabla_{\alpha}\phi\left(\nabla_{\mu}\nabla^{\alpha}\phi\right) = \mathcal{L}_{\rm m} Q\,,
\end{equation}
where $\nabla_{\mu}$ represents the covariant derivative, $\square \phi=g^{\mu \alpha} \nabla_{\mu} \nabla_{\alpha} \phi$, being $\square$ the d'Alembert operator and the coupling term $Q$ can then be expressed as
\begin{align}\label{coupling}
 Q = &\, -\frac{f_{,\phi}}{f} - \frac{f_{,X}}{f}\left(\square \phi+\partial^{\mu} \phi \frac{\nabla_{\mu} \mathcal{L}_{\rm m}}{\mathcal{L}_{\rm m}} +\frac{f_{,X}}{f} A + 2 \frac{f_{,\phi}}{f} X\right) + 2 \frac{f_{, X \phi}}{f} X+\frac{f_{, X X}}{f} A\,,
\end{align}
with $A=\nabla^{\mu} \phi \nabla_{\alpha} \phi \left(\nabla_{\mu} \nabla^{\alpha}\phi\right)$.

The conservation of the total energy-momentum tensor follows from the contracted Bianchi identities
\begin{equation}
\nabla_{\mu}G^{\mu}_{\nu} = 0 \quad \Rightarrow \quad \nabla_{\mu}\left( T^{(\phi)\,\mu}_{\quad\,\,\,\nu}+T^{{\rm (m)}\,\mu}_{\quad\,\,\,\,\,\nu} \right) = 0\,.
\end{equation}
Nevertheless, each component is not independently conserved due to the energy flow stemming from the interaction term $f$. The conservation relations are found to be
\begin{align}
\nabla_{\mu} T^{(\phi)\,\mu}_{\quad\,\,\,\nu} = &\, \mathcal{L}_{\rm m} Q\,\nabla_{\nu}\phi\,, \label{consphi} \\
\nabla_{\mu}T^{(m)\,\mu}_{\quad\,\,\,\,\,\nu} = &\, -\mathcal{L}_{\rm m} Q\,\nabla_{\nu}\phi\,,\label{consm}
\end{align}
with $Q$ given by Eq.~\eqref{coupling}.

Note that all the equations derived so far are completely independent of the choices of $g_{\mu\nu}$ and $f$, and thus valid on any background.

It is customary in the literature to impose the couplings on the conservation relations by fixing the term on the RHS of Eqs.~\eqref{consphi} and \eqref{consm} \cite{Baldi2012,Amendola2001,Olivares2006,Chen2008,Quercellini2008,Szydlowski2016,Arevalo2012,Nunes2001}. Coupled models dealing with noncanonical scalar fields were explored in \cite{Gumjudpai:2005ry,Chiba2014,Shahalam2017}, including interactions with nonlinear terms on $\dot{\phi}$ \cite{Das2015}, where a dot denotes derivative with respect to cosmic time, i.e. $\dot{\phi} \equiv \partial \phi/\partial t$. For the kinetic coupled dark energy, however, the coupling is imposed in the action through the choice of $f$. The conservation relations for the different interacting species appear \emph{a posteriori} from this choice. 

In this work, we accomplish the scaling regime by means of a ``fifth-force" acting on DM particles, generated by a tachyon field. It has already been shown in the literature \cite{Tamanini:2015iia} that an effective field theory formulation of our phenomenological interaction can be established at the level of the action which results in being able to build models, in a fully covariant manner, that are theoretically viable. Therefore, the propagation of unphysical modes on large scales is evaded \cite{Valiviita:2008iv}.

Such a successful approach can be achieved by inserting a coupling of the form $f(\phi)$ multiplying the CDM Lagrangian in the total action. The cases for a canonical and a tachyon scalar field were studied in \cite{Koivisto2005} and \cite{Farajollahi2011}, respectively. In \cite{Barros:2019rdv}, this formulation was generalised to include an interaction between the matter sector and the kinetic term of the scalar, through the form $f(\phi\,,X)$. More recently, several extensions and applications of this model have appeared in the literature. These include the study of an Horndesky Lagrangian for the scalar field using the Schutz-Sorkin action for matter \cite{Kase2020}, the analysis of perturbations and their corresponding stability \cite{Kase2020a}, the general form for the Lagrangian which allows for scaling solutions and an interacting Multi-Proca vector dark energy model \cite{Gomez2021}.

It is worth noting that recent developments \cite{Dunsby:2023qpb} have shown that a tachyon-like model appears among the fewest tachyonic fields that are observationally viable. It would be very interesting to compare this unified dark matter model with ours in a future work on the cosmological constraints of the kinetic coupled tachyon. Furthermore, since it is in principle possible to express a coupled dark energy model as a unified dark matter one, similarities and differences between the two models could then be explored. This would allow us to shed some light on the possibility of establishing tachyon-inspired models as strong alternatives to the standard cosmological scenario.

\subsection{A particular case}
\label{specificcases}

In case we neglect the coupling in the total action, Eq.~\eqref{actiongral}, that is, setting $f=1$, we recover the $k$-essence models thoroughly examined in the literature \cite{Armendariz-Picon:2000ulo, Gonzalez-Diaz:2003bwh,Scherrer:2004au,Babichev:2007dw,Chimento:2003ta,dePutter:2007ny}.
Furthermore, we can consider in this theory a tachyon scalar field,
\begin{equation}
\label{tachyon}
P(\phi,X) = -V(\phi)\, \sqrt{1-2X}\,,
\end{equation}
where $X$ is the kinetic term and $V(\phi)$ is a general self-coupling potential. Here the tachyon plays the role of dark energy and, obviously, we need $1-2X \geq 0$ in order to have a Lagrangian that is real valued. In this case, setting ${f}\equiv{f}(\phi)$ gives rise to the coupled tachyon model analysed in \cite{Farajollahi2011}.

If we now assume that our matter species is pressureless, the Lagrangian density can be written as \cite{Minazzoli:2012md,Harko:2010zi},
\begin{equation}\label{matterlagrangian}
\mathcal{L}_{\rm m}=T^{{\rm (m)} \alpha}{ }_{\alpha} = -\rho_{\rm m}\,,
\end{equation}
which results in a pressureless perfect fluid form for the energy-momentum tensor,
\begin{equation}\label{emmattertensor}
T^{\rm (m)}_{\mu\nu} = \rho_{\rm m} u_{\mu} u_{\nu}\,,
\end{equation}
where $u_{\mu}$ is the four-velocity vector, defined as $u_{\mu} u^{\mu}=-1$.

\section{Kinetic coupled tachyon}
\label{KCT}
In the previous section we have introduced the underlying formalism. Now we proceed to solve the field equations for the case of the tachyon scalar field with a kinetic coupling. In particular, we assume that the accelerated expansion of the Universe is driven by a tachyon scalar field, with $P(\phi,X) = -V(\phi)\, \sqrt{1-2X}$, kinetically coupled to a cold (pressureless) dark matter component whose Lagrangian density is given by $\mathcal{L}_{\rm m}=-\rho_{\rm m}$, with $\rho_{\rm m}$ being its energy density. Since couplings to standard model fields, such as radiation and baryon, are tightly constrained by observational data \cite{Anderson:2017phb,Carroll:1998zi,Devi:2011zz,Bertolami:2013qaa,Faraoni:2008ke}, it is customary to couple the scalar degree of freedom to dark matter, whose Lagrangian form is still unknown and that we assume it can be expressed through Eq.~\eqref{matterlagrangian}.

Thus, the total action is
\begin{equation}\label{kineticaction}
\mathcal{S}=\int \mathrm{d}^4 x \sqrt{-g} \left[\frac{M_{\rm Pl}^2}{2} R \overbrace{-V(\phi)\, \sqrt{1-2X}}^{=\mathcal{L}_{\phi}}+\underbrace{f(X) \tilde{\mathcal{L}}_{\rm m}}_{=\mathcal{L}_{\rm m}=-\rho_{\rm m}}\right] \,.
\end{equation}

We further stand upon a flat Friedmann-Lema\^itre-Robertson-Walker (FRLW) spacetime, with line element,
\begin{equation}
\label{metric}
\mathrm{d}s^2 = -\mathrm{d}t^2 +a(t)^2 \delta_{ij}\mathrm{d}x^i\mathrm{d}x^j\,,
\end{equation}
where $a(t)$ is the scale factor of the Universe as a function of cosmic time $t$.

Our choice for the potential is the following,
\begin{equation}
\label{potential}
V(\phi) =  \frac{V_0^2}{\phi^2}\,, \ \ \text{for}\  \ \lambda \neq 0  \,,
\end{equation}
where  $V_0\equiv 2M^2M_{\rm Pl}/\lambda$ is a mass scale associated with the scalar potential and $\lambda= - M^2M_{\rm Pl} V_{,\phi}/V^{3/2}$ stands for the stiffness of the potential, considered here as a dimensionless constant and a free parameter of the model.

Finally, we close the system by specifying a coupling function that depends only on the kinetic term,
\begin{equation}\label{couplingfunction}
f = (1-2X)^{\alpha/2}\,,
\end{equation}
where now $X=\frac{\dot{\phi}^2}{2M^4}$ and $\alpha$ is a constant that encapsulates the strength of the kinetic interaction. We recover the standard uncoupled tachyon model \cite{Bahamonde:2017ize} for $\alpha=0$. The factor $M^{4}$ through $X$ in Eq.~\eqref{couplingfunction} and in the definitions of $V_0$ and $\lambda$ simply renders the function $f$ dimensionless in the total action Eq.~\eqref{actiongral} \footnote{Remember that, in natural units, $[M_{\rm Pl}] = M$, $[\mathcal{L}] = M^{4}$, $[H] = M$, and for tachyon models $[\phi]=M$.}.

The tachyon field's energy density and pressure are defined as
\begin{equation}\label{energydensity}
\rho_{\phi}=-T^{(\phi) 0}{ }_{0}=\frac{V(\phi)}{\sqrt{1-\frac{\dot{\phi}^2}{M^4}}}\,,
\end{equation}
and
\begin{equation}\label{pressure}
p_{\phi} = \frac{1}{3}T^{(\phi) i}{ }_{i} = -V(\phi)\sqrt{1-\frac{\dot{\phi}^2}{M^4}}\,,
\end{equation}
respectively. The equation of state parameter is defined as
\begin{equation}\label{eos}
 w_{\phi} = \frac{p_{\phi}}{\rho_{\phi}} = -1 + \frac{\dot{\phi}^2}{M^4} \,,
\end{equation}
where the assumption $1-2X\geq0$ now translates into $\frac{\dot{\phi}^2}{M^4}\leq 1$. That is, we cannot have a phantom behaviour \cite{Amendola:2015ksp} since this requires $w_{\phi}<-1$. 

The tachyon model is a particular case of $k$-essence, although with different dynamics. Therefore, the same stability
conditions for $k$-essence models (see Ref.~\cite{Amendola:2015ksp}) apply also to the tachyon and these establish that the tachyon DE
model does not suffer from neither quantum instabilities nor superluminal
propagation.

Consideration of the $\nu=0$ component of Eqs.~\eqref{consphi} and \eqref{consm} yields the coupled equation of motion
\begin{align}\label{motionphi}
\ddot{\phi} & + \left(1-\frac{\dot{\phi}^2}{M^4} \right) \left(3H{\dot{\phi}}+M^4\frac{V_{,\phi}}{V}\right) 
= M^2\rho_{\rm m}\frac{Q}{V}\,\left(1-\frac{\dot{\phi}^2}{M^4} \right)^{3/2}\,,
\end{align}
and the conservation equations for the field and the matter components
\begin{align}
\dot{\rho}_{\phi} + 3H\rho_{\phi}(1+ w_{\phi}) = &\, \rho_{\rm m} Q\,\frac{\dot{\phi}}{M^2}\,, \label{continuityphi}\\
\dot{\rho}_{\rm m} + 3H \rho_{\rm m} = &\, -\dot{\phi}\, \rho_{\rm m} \, Q\,,\label{continuitymatter}
\end{align}
with $H\equiv \dot{a}/a$ being the Hubble function, $V_{,\phi} = \mathrm{d}V / \mathrm{d}\phi$, and from Eqs.~\eqref{coupling}, \eqref{matterlagrangian}, \eqref{couplingfunction}, \eqref{motionphi}, and \eqref{continuitymatter} we arrive at
\begin{equation}\label{coupling1}
Q = \frac{\alpha\left(1+ \frac{\dot{\phi}^2}{M^4}\right) \left(3H\frac{\dot{\phi}}{M^2} -\frac{\lambda\sqrt{V}}{M_{\rm Pl}}\right)}{\left[ \left(1-\frac{\dot{\phi}^2}{M^4}\right) + \alpha\left(1+ \frac{\dot{\phi}^2}{M^4}\right) \frac{ \rho_{\rm m}}{V}\, \sqrt{1-\frac{\dot{\phi}^2}{M^4}} - \alpha \frac{\dot{\phi}^2}{M^4}\right]}\,.
\end{equation}

The cosmic evolution is determined by the Friedmann equations, first the $00-$component of the field equations, Eq.~\eqref{EEs}, describing the expansion rate of the Universe,
\begin{equation}\label{eqfriedmann}
3H^2 M^2_{\rm Pl} =  \rho_{\phi}+\rho_{\rm m}\,,
\end{equation}
and second, the rate of change of the Hubble parameter,
\begin{equation}\label{secondfriedmann}
\dot{H} = -\frac{\rho_{\phi} + p_{\phi}+\rho_{\rm m}}{2M^2_{\rm Pl}}\,.
\end{equation}

\section{Dynamical system}
\label{DS}
We now proceed to study the evolution of the Universe for the kinetic coupled tachyon model and reduce the system of Eqs.~\eqref{motionphi}, \eqref{continuityphi}, \eqref{continuitymatter}, \eqref{eqfriedmann}, and \eqref{secondfriedmann} to a set of first order autonomous differential equations. In order to do so, we define the following dimensionless variables, generalising the ones already introduced in \cite{Copeland:2004hq},
\begin{equation}\label{variables}
x \equiv \frac{\dot{\phi}}{M^2}\,, \quad y \equiv \frac{\sqrt{V}}{\sqrt{3}HM_{\rm Pl}}\,, \quad z \equiv \frac{\sqrt{\rho_{\rm m}}}{\sqrt{3}HM_{\rm Pl}}\,. 
\end{equation}
As is well known in the literature for tachyon dark energy \cite{Aguirregabiria:2004xd,Copeland:2004hq}, the dynamical system with the inverse square potential yields the simplest closed system of autonomous equations. This is in stark contrast with the quintessence case, which is characterised by an exponential potential. 
Armed with these variables, we are now in a position to define the density parameter for the tachyon field and matter
\begin{align}
\Omega_{\phi} = &\, \frac{y^2}{\sqrt{1-x^2}}\,, \label{omegaphi}\\
\Omega_{\rm m} = &\, z^2\,. \label{omegam}
\end{align}

From Eqs.~\eqref{omegaphi} and \eqref{omegam} we can express the Friedmann constraint in terms of the dimensionless variables
\begin{equation}
\Omega_{\phi} + \Omega_{\rm m} =1 \Longrightarrow\frac{y^2}{\sqrt{1-x^2}} + z^2=1\,,
\label{friedcons}
\end{equation}
which allows us to replace $z$ in terms of $x$ and $y$, thus reducing the dimensionality of the dynamical system. 
The coupling term, Eq.~\eqref{coupling}, can also be recast in terms of the dynamical variables, yielding
\begin{equation}\label{interaction}
Q =  \frac{\alpha (1+ x^2) \left( 3x-\sqrt{3}\lambda y \right)H}{\left[ (1-x^{2}) + \alpha(1+ x^2) \frac{ z^2}{y^2}\,\left(1-x^2 \right)^{1/2} - \alpha x^2\right]}\,.
\end{equation}
Obviously, when $\alpha=0$ the system reduces to the uncoupled case. 
The system of autonomous equations obtained thanks to the variables defined in Eq.~\eqref{variables} is then 
\begin{align}
 x^{\prime} = &\, \left(1-x^2\right) \left(\sqrt{3}\lambda y - 3x \right) 
 \times \left[\frac{(1-x^{2})y^2 - \alpha x^2y^2}{(1-x^{2})y^2+ \alpha(1+x^2)z^2\,\sqrt{1-x^2}-\alpha x^2y^2}\right]\,, \label{xl} \\
 y^{\prime} = &\, -\frac{1}{2}y\left(\sqrt{3}\lambda xy + 2\frac{H^{\prime}}{H}\right)\,,\label{yl} \\
 z^{\prime} = &\, -\frac{1}{2}z\left[3 + 2\frac{H^{\prime}}{H} + 
 \frac{\alpha x\left(1+x^2\right)\left(3x-\sqrt{3}\lambda y\right)}{1-x^2 + \alpha\left(1+x^2\right)\frac{z^2}{y^2}\sqrt{1-x^2}-\alpha x^2} \right]\,,\label{zl} 
\end{align}
where a prime denotes the derivative with respect to the number of e-folds, $N\equiv\ln a$, and we have used 
\begin{equation}
\frac{H'}{H}=-\frac{3}{2}\left( 1+ w_{\rm eff} \right)\,,
\end{equation}
with
\begin{equation}
\label{eqweff}
w_{\rm eff}= -y^2\sqrt{1-x^2}\,,
\end{equation}
being the effective equation of state parameter. We require $w_{\rm eff}<-1/3$ at present since we are undergoing a period of accelerated expansion.

When we use Eq.~\eqref{friedcons} to replace $z$ in Eqs.~\eqref{xl} -- \eqref{zl} to reduce the dimensionality of the system, we end up with 
\begin{align}
 x^{\prime} = &\, \left(1-x^2\right) \left(\sqrt{3}\lambda y - 3x \right) 
 \times \left[\frac{(1-x^{2})y^2 - \alpha x^2y^2}{(1-x^{2})y^2+ \alpha(1+x^2)\, (\sqrt{1-x^2}-y^2)-\alpha x^2y^2}\right]\,, \label{xl1} \\
 y^{\prime} = &\, -\frac{1}{2}y\left[\sqrt{3}\lambda xy -3\left(1-y^2\sqrt{1-x^2}\right)\right]\,. \label{yl1} 
\end{align}

The equation of state parameter for the tachyon field is defined in terms of the dimensionless variables as
\begin{equation}
\label{wphi}
w_{\phi} = \frac{p_{\phi}}{\rho_{\phi}}  = -1 + x^2 \,.
\end{equation}

On a final note, notice that the coupling function, Eq.~\eqref{couplingfunction}, diverges if $\alpha<0$ and $\frac{\dot{\phi}^2}{M^4}=1$. On the other hand, the coupling term, Eq.~\eqref{interaction}, diverges for negative values of $\alpha$ along the values 
\begin{equation}
y^2 = \frac{\alpha \left(1+x^{2}\right)\sqrt{1-x^{2}}}{\alpha \left(1+2x^{2})-(1-x^{2}\right)} \,,
\end{equation}
and, as a result, the phase-space is not properly defined. Therefore, from now onwards, we shall consider that $\alpha\geqslant 0$.

We would like to point out that the choice of variables that allows us to write the evolution equations as an autonomous system is not unique. It is customary in cosmology to use for that purpose the set of dimensionless variables introduced in \cite{Copeland:1997et} (also known as expansion-normalised variables \citep{Wainwright_Ellis2005}). We have modified this procedure for the case of the tachyon. This does not alter the dynamics of the system, and it is more convenient than other choices for the variables that could render the analysis unnecessarily involved.

\subsection{\label{sec:ps} Phase space and invariant sets}
Given that $ 0 \leq \Omega_{\phi} \leq 1$, the allowed range for $x$ and $y$ is
\begin{equation}
0\leq x^2 + y^4 \leq 1\,.
\label{rest}
\end{equation} 
This is our physical phase-space, in other words, the invariant set which contains all of the orbits that are physically relevant. Notice that both $x$ and $y$ are finite in the range $0\leq x^2  \leq1$ and $0\leq  y \leq1$.

By inspecting the system of Eqs.~\eqref{xl} -- \eqref{zl}, we immediately see that the dynamical system is symmetric under the time reversal $t\mapsto -t$. In addition, the dynamical system is also invariant under the transformations: $(x,y) \longmapsto (-x,-y)$, and $(x, \lambda) \longmapsto (-x,-\lambda)$. This means that it is enough to consider in our analysis the upper half disk $y\geqslant0$ and the region corresponding to non-negative values of $\lambda$. Moreover, we consider $H>0$, imposing, therefore, an expanding Universe.

\subsection{\label{sec:fp} Fixed points and phase-space analysis}

In order to find the fixed points of the autonomous system Eqs.~\eqref{xl} -- \eqref{yl}, we set $(x',y')=(0,0)$. The fixed points are shown in Table~\ref{tab:critpoints}, where, for simplicity, we have used
\begin{equation}
 y_{\rm d} = \sqrt{\frac{\sqrt{\lambda^4+36}-\lambda^2}{6}}\,,
\label{eq:yd}
\end{equation}
and
\begin{equation}
 y_{\rm s,\pm} = \frac{\sqrt{\alpha[12\sqrt{\alpha(1+\alpha)}+\lambda^2]}\pm\sqrt{\alpha}\lambda}{2\sqrt{3}\alpha}\,.
 \label{eq:ys}
\end{equation} 

These critical points depend on the two free parameters of our model, i.e. the kinetic coupling $\alpha$ and the stiffness of the potential $\lambda$. Given that the fixed points are all hyperbolic, their stability is amenable to be obtained through linear stability analysis.

We now proceed to investigate the existence and stability of the critical points of the autonomous equations Eqs.~\eqref{xl} -- \eqref{yl}, and whether they generate an accelerated expansion. 

The existence is explored with the condition $0\leq \Omega_{\phi} \leq 1$ that leads to Eq.~\eqref{rest}. This constraint also ensures that the energy density of the matter component is either zero or positive. 

For studying the stability, a linear perturbation around the fixed points is considered with a small perturbation $\delta x$
\begin{align*}
x & \to x + \delta x \,,\\
y & \to y + \delta y \,.
\end{align*}
Thus the linear autonomous equations Eqs.~\eqref{xl} -- \eqref{yl} take the form 
\begin{align}
\begin{pmatrix}
\delta x\\ \delta y
\end{pmatrix}^{'}= \mathcal{M} \begin{pmatrix}
\delta x\\ \delta y
\end{pmatrix}\,,
\label{eq:linearpert}
\end{align}
and the solution of the system would be
\begin{align*}
\Vec{\delta} = C_{1}\Vec{V_1}\,e^{\mu_1 N} + C_{2}\Vec{V_2}\,e^{\mu_2 N}\,,
\end{align*}
where $\mu_i$ are the eigenvalues of the system, $\Vec{V_i}$ the eigenvectors and $C_i$ integration constants. Thus, the eigenvalues of the matrix $\mathcal{M}$ in Eq.~\eqref{eq:linearpert} show the behaviour of the linear perturbations around the critical points. These eigenvalues and the corresponding stability states are given in Table~\ref{tab:stabilitycritpoints}.
The stability analysis is performed by inspecting the eigenvalues. If all the $\mu_i$ are negative, then all the perturbations decay, and we say that the critical point is stable. If all the eigenvalues are positive, then the perturbations grow and we have an unstable critical point. Finally, if one $\mu_i$ is positive and the other negative, then the perturbations grow along given directions and decay over others (along the directions of the respective eigenvectors), and we say that the point is a saddle.

\begin{table}[ht]
\caption{\label{tab:stabilitycritpoints} Eigenvalues of the matrix $\mathcal{M}$ and the stability implied by them for the four critical points in the kinetic coupled tachyon model.}
\centering
 \begin{tabular}{|c|c|c|c|}
  \hline
  \hline
  Point & $\mu_1$ & $\mu_2$ & Stability\\
  \hline
  (A) & $0$ & $\tfrac{3}{2}$ & Unstable \\
  \hline
  (B) & $-3y_{\rm d}^4$ & $-\tfrac{3}{2}(1+y_{\rm d}^4)$ & Stable \\
  \hline
  (C) & $-6\tfrac{\sqrt{\alpha(1+\alpha)}}{(2+\alpha)^2}y_{\rm s,+}^2$ & $-\tfrac{3}{2}\tfrac{\alpha y_{\rm s,+}^2+\sqrt{\alpha(1+\alpha)}}{\sqrt{\alpha(1+\alpha)}}$ & Stable \\
  \hline
  (D) & $-6\tfrac{\sqrt{\alpha(1+\alpha)}}{(2+\alpha)^2}y_{\rm s,-}^2$ & $-\tfrac{3}{2}\tfrac{\alpha y_{\rm s,-}^2+\sqrt{\alpha(1+\alpha)}}{\sqrt{\alpha(1+\alpha)}}$ & Stable \\
  \hline
  \hline
 \end{tabular}
\end{table}

\begin{table*}[ht]
 \caption{\label{tab:critpoints} Critical points for the dynamical system for coupled tachyon models.}
 \centering
 \adjustbox{margin=-1.8cm 0cm 0cm 0cm}{
 \begin{tabular}{|c|c|c|c|c|c|c|c|}
  \hline
  \hline
  Point & $x$ & $y$ & $\Omega_{\phi}$ & $w_{\phi}$ & $w_{\rm eff}$ & Existence & Acceleration \\
  \hline
  (A) & $x_{\rm a}$ & 0 & 0 & $-1+x_{\rm a}^2$ & 0 & $\forall \alpha,\lambda$ & No \\
  \hline
  (B) & $\frac{\sqrt{3}}{3}\lambda y_{\rm d}$ & $y_{\rm d}$ & 1 & $-1 + \frac{\lambda^2}{3}y_{\rm d}^2$ & $-1 + \frac{\lambda^2}{3}y_{\rm d}^2$ & $\forall \alpha,\lambda$ & $-12^{1/4}<\lambda<12^{1/4}$ \\
  \hline
  (C) & $-\frac{1}{\sqrt{1+\alpha}}$ & $y_{\rm s,+}$ & $\sqrt{\tfrac{1+\alpha}{\alpha}}y_{\rm s,+}^2$ & $-\tfrac{\alpha}{1+\alpha}$ & $-\sqrt{\tfrac{\alpha}{1+\alpha}}y_{\rm s,+}^2$ & $\lambda\le -\tfrac{\sqrt{3}}{\left[\alpha(1+\alpha)\right]^{1/4}}$ & $-2\left[\alpha(1+\alpha)\right]^{1/4}<\lambda<0$ \\
  \hline
  (D) & $\frac{1}{\sqrt{1+\alpha}}$ & $y_{\rm s,-}$ & $\sqrt{\tfrac{1+\alpha}{\alpha}}y_{\rm s,-}^2$ & $-\tfrac{\alpha}{1+\alpha}$ & $-\sqrt{\tfrac{\alpha}{1+\alpha}}y_{\rm s,-}^2$ & $\lambda\ge \tfrac{\sqrt{3}}{\left[\alpha(1+\alpha)\right]^{1/4}}$ & $0<\lambda<2\left[\alpha(1+\alpha)\right]^{1/4}$ \\
  \hline
  \hline
 \end{tabular}
}
\end{table*}

Lastly, a Universe experiencing an accelerated expansion has an effective equation of state which is smaller than $-1/3$. In our system, the effective equation of state ($w_{\rm eff}$) reads as $-y^2\sqrt{1-x^2}$. Therefore, we look for the condition
\begin{equation}
w_{\rm eff}= -y^2\sqrt{1-x^2}<-1/3\,.
\end{equation}

Before starting to study each point, there are some physical constraints we apply. First, all critical points must be real so we eliminate critical points with complex values. Then, by definition, $y$ must be greater than 0, because it is depending on the square root of the tachyonic field potential, so we only consider the upper half of x-y coordinate. 

With all these conditions we have four critical points as it can be seen in Table \ref{tab:critpoints}. There is also an additional point which we do not add to the table, with coordinates $(1,\sqrt{3}/\lambda)$. For that point, $\Omega_{\phi}\to\infty$ as noticed in \cite{Bahamonde:2017ize} when $w_{\rm m}=0$. We shall, therefore, not consider it further.

\paragraph*{\textbf{Point A:}} $(x,y) \rightarrow (x_{\rm a},\,0)$\\
Here $x_{\rm a}$ is an arbitrary x-coordinate. It describes a line of hyperbolic
critical points (critical line). This critical point corresponds to a matter dominated solution with $\Omega_{\rm m} = 1$ ($\Omega_{\phi} = 0$), and it exists for all values of $\alpha$ and $\lambda$. The tachyon scalar field is negligible around this point. This point is then unable to create an accelerated expansion. This can also be seen from the zero effective equation of state ($w_{\rm eff} = 0$). The stability requirements identify it as a saddle point and it is never stable. It attracts the orbits towards the origin of the phase-space when the coupling is absent and repels them towards the $y$-axis when the coupling is present, i.e. when $\alpha \neq 0$. In other words, when the kinetic coupling is present, the critical point is altered and becomes a repeller. 

\begin{figure*}[ht]
    \centering
    {\hspace*{-4.9cm}\includegraphics[width=1.7\textwidth]{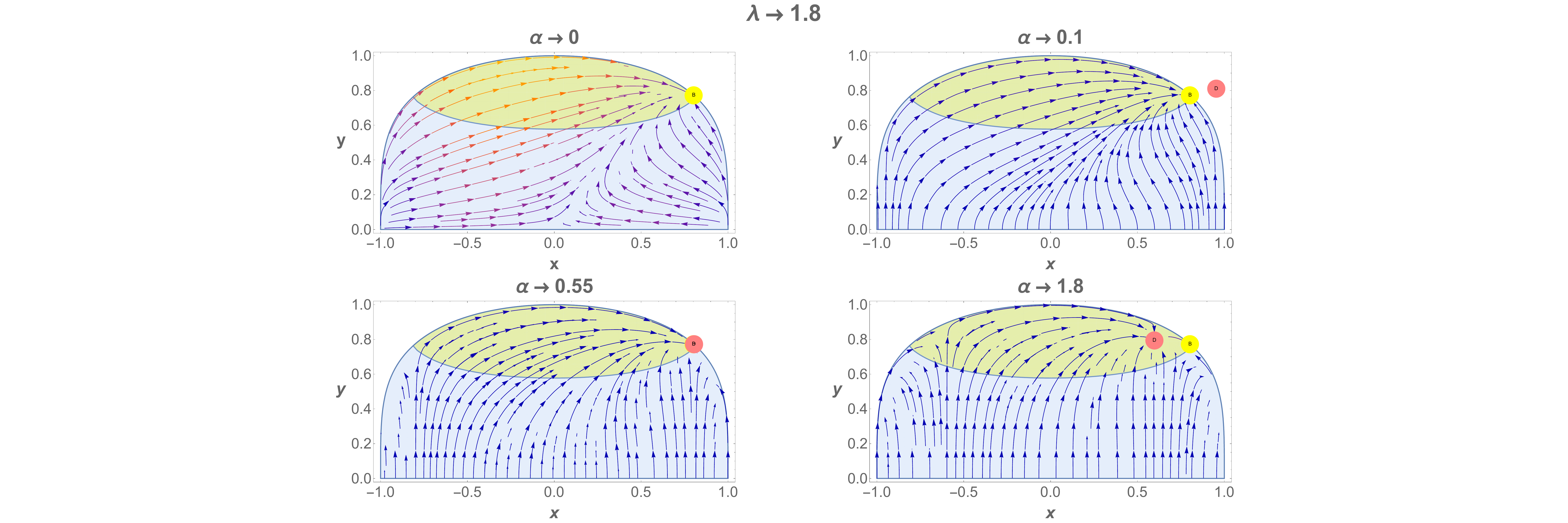}}
    \caption{The phase-spaces of the autonomous system defined in Eqs.~(\ref{xl}) -- (\ref{zl}) are plotted with the stiffness of the potential ($\lambda$) equal to 1.8. They are depicted for four different values of the coupling $\alpha$, ranging from 0 to 1.8, which are noted above each plot. The blue region where orbits lie denotes the existence ($0\leq y^4 +x^2 \leq 1$). The yellow region, on the other hand, represents the parameter region where the Universe experiences an accelerated expansion. There are two critical points visible on the phase spaces and they are painted in blue and pink and labelled as B and D, respectively. Notice that when $\alpha = 0.55$, both critical points merge.}
    \label{fig:phase_space}
\end{figure*}

\paragraph*{\textbf{Point B:}}$(x,y) \rightarrow (\frac{\sqrt{3}}{3}\lambda y_{\rm d},\,y_{\rm d})$\\
This critical point was also found in the uncoupled case \cite{Copeland:2004hq}, and represents a tachyonic-field dominated solution, with $\Omega_{\phi} = 1$ and $\Omega_{\rm m} = 0$, with the variable $y_{\rm d}$ being defined in Eq.~\eqref{eq:yd}. This fixed point depends explicitly on $\lambda$, it exists for all $\lambda$ and $\alpha$ and it is stable. As it can be seen in Fig.~\ref{fig:phase_space}, it is always located on the boundary $x^2+y^4=1$. The equation of state is given by $w_{\phi} = w_{\rm eff} = -1 + \frac{\lambda^2}{3}y_{\rm d}^2$. For $\lambda^4<12$, it features accelerated expansion, irrespective of the value of the coupling $\alpha$, being not only an attractor solution but the only attractor solution (a global attractor) of the system for the uncoupled case. It is located inside the yellow region in Fig.~\ref{fig:phase_space}, where the Universe goes through accelerated
expansion. Notice that when $\lambda \rightarrow 0$, it displays a cosmological constant-like behaviour.

We can use this uncoupled case to achieve a transition from the dark matter era and the dark energy dominated era. Nevertheless, due to the nature of the fixed point, the Universe will become completely dark-energy dominated, that is, it will expand forever. Notice, however, that if we want to replicate the current observed matter content, we are forced to choose specific initial conditions such that the attractor is still not reached today. This particular choice of initial condition leads to the fine-tuning problem.

\paragraph*{\textbf{Point C:}}$(x,y) \rightarrow (-\frac{1}{\sqrt{1+\alpha}},\,y_{\rm s_+})$\\
This fixed point emerges due to the presence of coupling ($\alpha \neq 0$). It gives rise to a scaling solution, with $\Omega_{\phi} = \sqrt{\tfrac{1+\alpha}{\alpha}}y_{\rm s_+}^2$, where $y_{\rm s_+}$ is defined in Eq.~\eqref{eq:ys}. This critical point generates acceleration when 
\begin{equation}
-2\left[\alpha(1+\alpha)\right]^{1/4}<\lambda<0 \,,
\end{equation}
and it exists only for 
\begin{equation}
\lambda\le -\tfrac{\sqrt{3}}{\left[\alpha(1+\alpha)\right]^{1/4}} \,.
\end{equation}

Since we consider that the mass scale, $V_0$, defined in Eq.~\eqref{potential}, is positive, this leads to $ \lambda > 0$. Therefore, this point does not exist and never generates acceleration in the framework of this paper. However, this point is stable for every $\lambda$ and $\alpha$.\\

\paragraph*{\textbf{Point D:}}$(x,y) \rightarrow (\frac{1}{\sqrt{1+\alpha}},\,y_{\rm s_-})$\\
Just as it happens for the critical point C, this novel solution only emerges when the kinetic coupling is present ($\alpha \neq 0$).
It corresponds to a kinetic scaling fixed point, given that the cosmological parameters $w_{\rm eff}$ and $\Omega_{\phi}$ depend on both $\lambda$ and $\alpha$, with $\Omega_{\phi} = \sqrt{\tfrac{1+\alpha}{\alpha}}y_{\rm s_-}^2$, where, again, $y_{\rm s_-}$ is defined in Eq.~\eqref{eq:ys}. In the absence of coupling, the tachyonic scalar field density evolves in the same way as the matter around this point and when $\alpha \rightarrow \infty$ the scalar field density reaches a constant value. Contrary to point C, it exists for positive values of $\lambda$ and it is thus a valid fixed point for our analysis. The condition for its existence is given by
\begin{equation}\label{existenceD}
\lambda\ge \tfrac{\sqrt{3}}{\left[\alpha(1+\alpha)\right]^{1/4}}\,,
\end{equation}
and acceleration is generated when 
\begin{equation}
0<\lambda<2\left[\alpha(1+\alpha)\right]^{1/4}\,.
\end{equation}

The conditions stated above imply that this fixed point is visible in the phase-space (see Fig.~\ref{fig:phase_space}) for a stronger coupling. In the same figure, it can also be seen that as soon as the coupling is strong enough, this critical point plays the role of an attractor solution. In particular, solving Eq.~\eqref{existenceD} for the equal sign, the points B and D coincide and merge, becoming the global attractor of the system. For the value of $\lambda=1.8$, the solution is $\alpha= 0.55$, and it is depicted in Fig.~\ref{fig:phase_space}. This is again a stable point for every $\lambda$ and $\alpha$. All of the orbits depicted in Fig.~\ref{fig:phase_space} correlate to solutions connecting the fixed point (A) to either
(B) or, when it happens to exist, (D), except for the orbit connecting (B) to (D).

We acknowledge the fact that for a fixed value of the parameters involved in $w_{\rm eff}$, we can obtain an accelerating Universe thanks to this scaling solution. This is an attractive feature because the scalar field reminds hidden during the early Universe era, with an energy density that can be sizeable but nevertheless negligible at present. This would generate a natural mechanism (independent of the initial conditions), to explain the fact that the energy densities of DE and DM are of the same order today. As a matter of fact, this holds not only near the present but throughout a much larger period of the history of our Universe. Note that this effect is impossible to achieve with $\Lambda$ as its energy density is always constant throughout time. This scaling solution alleviates therefore the coincidence problem, which is certainly appealing, given that the observed present value of $\Omega_{\phi} \thickapprox 0.7$ \cite{Planck:2018vyg} can be obtained irrespective of how we set the initial conditions. Eventually, the final state of the Universe will be attained when $\Omega_{\phi} = 1$.

\begin{figure}[ht]
    \centering
    \includegraphics[width=0.8\textwidth]{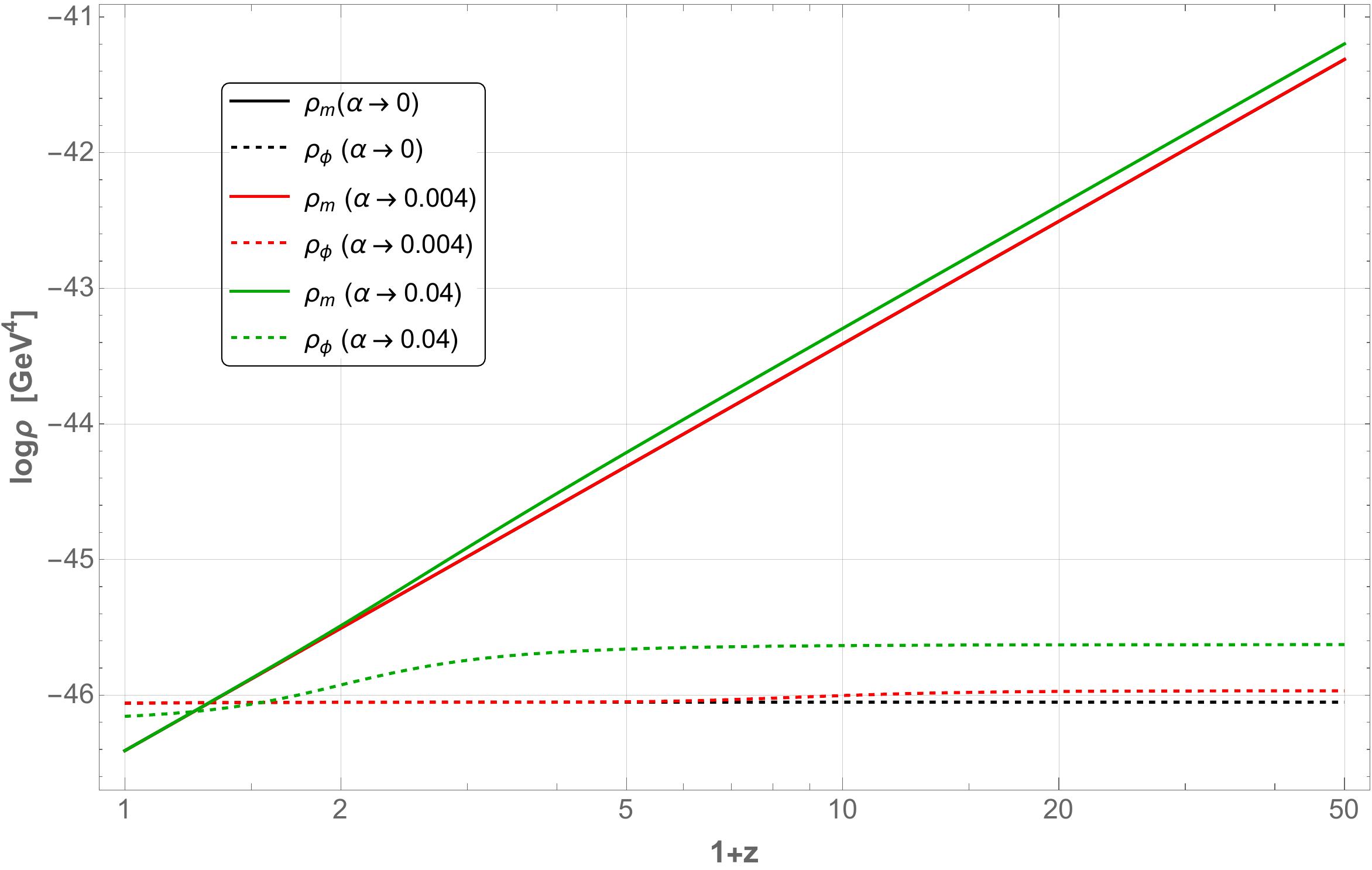}
    \caption{Evolution of the energy densities for the tachyonic scalar field ($\rho_{\phi}$) and matter ($\rho_{\rm m}$) for the solutions of Eqs.~\eqref{xl} -- \eqref{zl} with $\lambda=0.3$ and three different choices for $\alpha$.}
    \label{fig:relative_density}
\end{figure}

\begin{figure}[ht]
    \centering
    \includegraphics[width=0.8\textwidth]{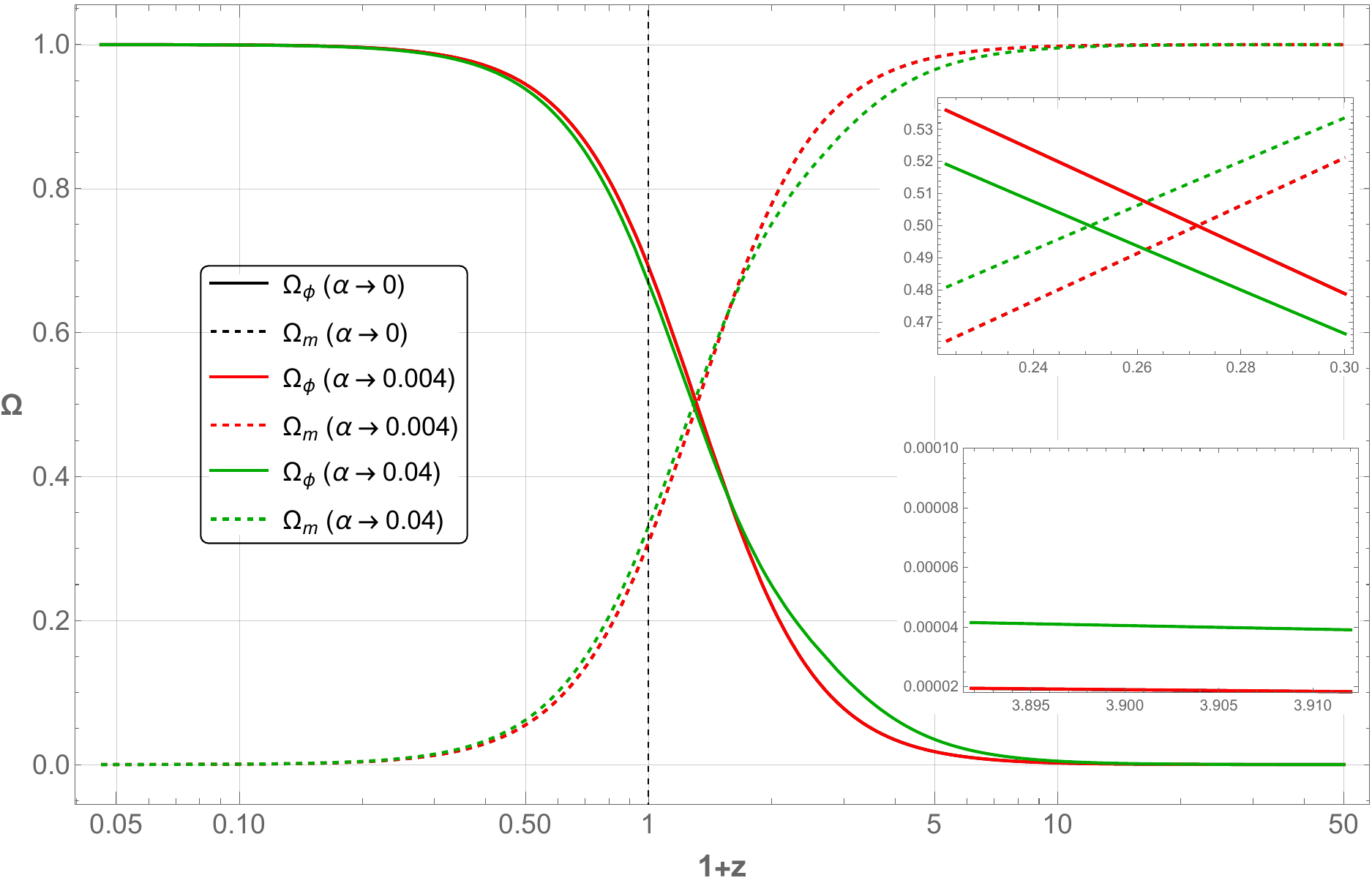}
    \caption{Evolution of the relative energy densities for the tachyonic scalar field ($\Omega_{\phi}$) and matter ($\Omega_{\rm m}$) for $\lambda=0.3$ and three different values of $\alpha$. Note that the black curves relative to $\alpha=0$ are indistinguishable from the red curves.}
    \label{fig:relative_omegas}
\end{figure}

\begin{figure}[ht]
    \centering
    \includegraphics[width=0.8\textwidth]{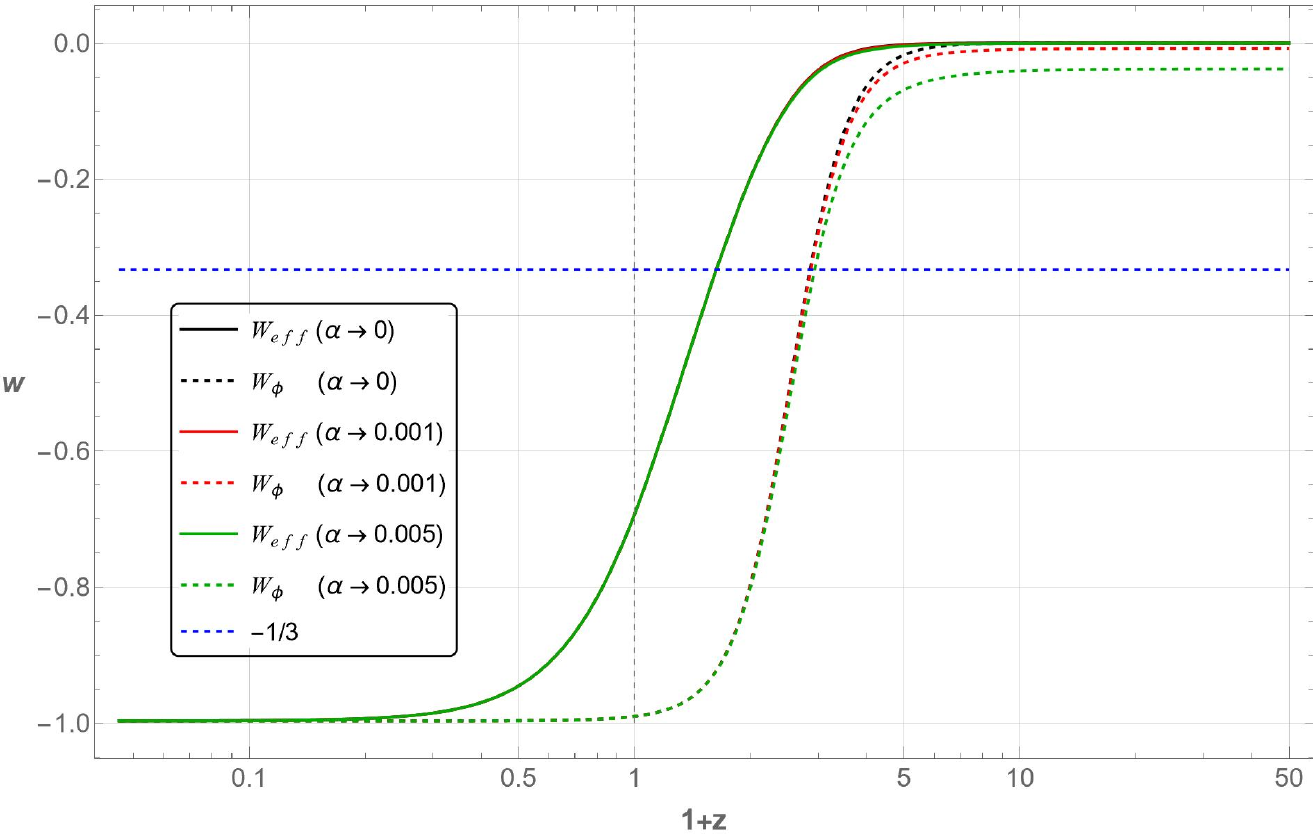}
    \caption{Evolution of the equation of state of the tachyonic scalar field ($w_{\phi}$) and the effective equation of state ($w_{\rm eff}$) for $\lambda=0.1$ and three different values of $\alpha$.}
    \label{fig:eos}
\end{figure}

Notice that near this attractor, we always have $\dot{\phi} > 0$. Furthermore, the sign of the coupling $Q$, defined in Eq.~\eqref{interaction}, is the same as the sign of the kinetic coupling $\alpha$, determining the direction of the energy flow between DM and DE in Eqs.~\eqref{continuityphi} and \eqref{continuitymatter}. Since for us $\alpha > 0$, the matter sector is sourcing the tachyon field. This implies that given the existence of the scaling solution, near the attractor, there will always be an energy exchange from the matter component to the tachyon field $\phi$.

The initial conditions were computed by setting $\Omega_{\phi}^0\approx 0.7$, according to the value indicated by Planck \cite{Planck:2018vyg}.
This was followed by a root finding applied to a search in the parameter space in order to find the right initial conditions, such that the initial value for $y$ always lies in the vicinity of the critical point (D) so that the scaling for higher alpha and lambda values is guaranteed.

In Fig.~\ref{fig:relative_density} we show the energy density of the constituents of the two-component universe, tachyonic scalar field ($\rho_{\phi}$) and pressureless matter ($\rho_{\rm m}$), for the solutions of Eqs.~\eqref{xl} -- \eqref{zl} when $\lambda = 0.3$ and three different values of $\alpha$. We observe that as $\alpha$ increases, the contribution of the tachyon scalar field in the early universe increases. We do not see the effect of the scaling regime when the attractor (D) appears. This is expected considering the stiffness of the potential and the coupling strength used. As we already discussed, the scaling solution appears as a global attractor of the system when the coupling to the matter is increased. We also show that the energy transfer between the matter and the field starts earlier in the history of the Universe for larger couplings. Likewise, in the scenarios with a smaller coupling, it takes less time for the field to converge to the full domination of the component ($\Omega = 1$).

In Fig.~\ref{fig:relative_omegas}, the evolution of the relative energy densities of the tachyon scalar field and the matter content, $\Omega_{\phi}$ and $\Omega_{\rm m}$, is plotted for $\lambda = 0.3$ and three illustrative values of $\alpha$. On close inspection of the figure, we can clearly see that the transition from a matter dominated Universe to a DE dominated one occurs at a later stage for larger values of the coupling. On the other hand, for $\alpha= 0.04$ we see that the value of $\Omega_{\phi}$ seems to indicate a small amount of early DE whose presence may have an effect on the position of the CMB peaks \cite{Wetterich:2004pv,Planck:2015bue}. Nevertheless, for smaller values of the coupling, such as $\alpha= 0.004$, this effect becomes less important during the matter dominated epoch. 
Incidentally, given that the evolution of the cosmic fluids in models with a scalar field is modified with respect to the one in the standard model, this will have a direct impact on the rate at which the Universe expands (through the Friedmann equation). Therefore, the theoretical predictions for $H_{0}$ in such models will differ from the value found in $\Lambda$CDM, which may be of interest to deal with the present discrepancy between early and local measurements of $H_{0}$. In this regard, models with early dark energy may present a larger expansion rate at early times \cite{Karwal:2016vyq} and, as a result, they can alleviate the $H_0$ tension \cite{Poulin:2018cxd}.

Fig.~\ref{fig:eos} illustrates how the equation of state for the tachyon field ($w_{\phi}$) and the effective equation of state ($w_{\rm eff}$) evolve for a fixed $\lambda = 0.1$ and different values of the coupling $\alpha$. In the absence of the coupling, that is, when $\alpha = 0$, the evolution of the path moves towards the attractor B, the global attractor in the uncoupled scenario, since no scaling regime is taking place in this case. However, when the interaction is introduced, this triggers the appearance of a new critical point (D), which is a scaling solution and an attractor of the system. As a consequence, the trajectory is slowed, and this behaviour becomes more apparent with growing $\alpha$. We can also notice that this effect is virtually inactive in $w_{\rm eff}$, and, therefore, its trajectory is hardly modified in the presence of the coupling. This outcome can be attributed to the fact that the DE contribution at early times is insignificant for the parameters selected in Fig.~\ref{fig:eos}. Once we enter in the DE dominated era, the tachyon scalar field takes over and drives the acceleration ($w_{\rm eff}< -1/3$).

On a final note, as it was discussed in \cite{Copeland:2006wr}, the only feasible solution in the uncoupled case is the scalar field dominated fixed point (B). This is the only attractor of the system but as it can be seen in Table~\ref{tab:critpoints}, in order to generate accelerated expansion, we need $\lambda^4 <12$. This implies  a constraint on the mass scale of the potential $V_0$ such that $V_0 \gtrsim 1.1 M_{\rm Pl}$. In other words, the energy scale of the potential needs to be greater than the Planck mass, which is in stark contrast with what we would expect, given that General Relativity is supposed to break down at this scale. This manifest conundrum can be ameliorated thanks to the scaling solution (D). For this critical point, $\lambda$ can take any value between $0<\lambda<2\left[\alpha(1+\alpha)\right]^{1/4}$, that is, there is no longer  a constraint on the mass of the potential since the phenomenon is now moved to the scale of the kinetic coupling.

Recently, some popular dark energy models, characterised by viable scalar field potentials,  have been reviewed and their stability and the dynamical systems associated with them have been studied \cite{Carloni:2024rrk}. This was motivated by the renewed interest in evolving dark energy models, supported by the results coming from the DESI collaboration \cite{DESI:2024mwx}. However, although our choice for the potential is included in the set of possible dark energy potentials considered in \cite{Carloni:2024rrk}, we cannot establish a direct comparison with their results since a tachyon scalar field was not explored in their investigation. In the case of the tachyon, the typical potential is $V(\phi)=V_{0}/\cosh{(\beta\phi/2)}$ with $\beta$ being either $1$ or $\sqrt{2}$ \cite{Kutasov:2003er} but when $\phi$ is large this potential is too steep to maintain a late time accelerated expansion. On the other hand, it has been shown that the inverse square potential ($V(\phi) \varpropto \phi^{-2}$) can account for an expansion of the Universe at late times \cite{Padmanabhan:2002cp,Abramo:2003cp,Aguirregabiria:2004xd,Copeland:2004hq} and that is why it is our preferred choice. That said, if once our model is tested against the latest observational data we find that this potential is not suitable for the viability of our model, we may contemplate the case given by $V(\phi) = V_{0} e^{\frac{1}{2}m^{2}\phi^{2}}$, which it is also used in \cite{Carloni:2024rrk}, since it is also possible to have an accelerated expansion in this scenario \cite{Garousi:2004uf}.

\section{Conclusions}
\label{Concl}
In this paper we have studied a generalisation of interacting DE cosmological models where the kinetic term of a tachyon field couples to the matter sector at the level of the action. The action is presented in Sect.~\ref{model} along with the modified field equations and the conservation equations for both species.

We have obtained the cosmological equations in Sect.~\ref{KCT} where a tachyon field $\phi$ is coupled to CDM by means of a particular form of the kinetic coupling. We have taken an inverse squared potential, which is the simplest case related to the tachyon field $\phi$. This choice implies considering $\lambda$, the stiffness of the potential, as a constant. This model has been thoroughly examined in the literature for the uncoupled case. In the absence of the coupling, the attractor will always be the scalar field dominated fixed point (B) which presents an accelerated
expanding behaviour for $\lambda^{4} < 12$. However, from a phenomenological point of view, there is no reason to believe that the components of the dark sector cannot interact. The advantage of considering an interaction between DM and the DE field when studying dynamical systems is the possibility of having scaling solutions. This is precisely what we have found in our dynamical analysis study, where the kinetic coupling $\alpha$ gives rise to a viable fixed point not present in the uncoupled case, which is a stable attractor and that also provides accelerated expansion when $0<\lambda<2\left[\alpha(1+\alpha)\right]^{1/4}$. This fixed point (D) exhibits an early scaling regime, with only a tiny presence of DE during the matter domination epoch. This is followed by a period with accelerated expansion, with a late time attractor.

The scaling solution obtained can alleviate the coincidence problem, which is an attractive feature. We have also found that the transition from a matter dominated era to a DE era takes place earlier in the cosmological history of the Universe when the coupling is stronger.

Finally, it is relevant to realise that, unlike the standard coupled tachyon scenarios, in the model presented here the kinetic term of the scalar field is allowed to couple to the matter fluids \textit{a priori} in the action. The implication of this underlying theory is that this model encompasses a gamut of already known DE models, obtained when the different functions within the theory take particular values.

The study of the full linear behaviour and the growth of CDM density perturbations in the nonlinear regime for this model are currently underway.

\section*{Acknowledgements}
We gracefully thank Nelson J. Nunes for help with Mathematica. F.P. acknowledges partial support from the INFN grant InDark and from the Italian Ministry of University and Research (\textsc{mur}), PRIN 2022 `EXSKALIBUR – Euclid-Cross-SKA: Likelihood Inference Building for Universe's Research', Grant No.\ 20222BBYB9, CUP C53D2300131 0006, and from the European Union -- Next Generation EU. FP and ARF acknowledge support from the FCT project ``BEYLA -- BEYond LAmbda" with ref. number PTDC/FIS-AST/0054/2021. \"{O}TT acknowledges financial support from Coordena\c{c}\~{a}o de Aperfei\c{c}oamento de Pessoal de N\'{i}vel Superior (CAPES) for his PhD fellowship.

\appendix

\section{Kinetic coupled dark energy}
We start from the action in Eq.~\eqref{actiongral} and apply Hamilton’s principle. The dynamics of $\phi$ is then obtained when the action is minimised
\begin{equation}\label{action}
 \begin{split}
\delta \mathcal{S} = &\, \delta S_{\phi}+\delta S_{\rm m} \\ 
= &\, \int \mathrm{d}^4 x \sqrt{-g} \frac{\delta P(\phi, X)}{\delta \phi} \delta \phi 
 + 
 \int \mathrm{d}^4 x \sqrt{-g} \frac{\delta\left[f(\phi, X) \tilde{\mathcal{L}}_{\rm m}\right]}{\delta \phi} \delta \phi=0\,,
 \end{split}
\end{equation}
being $X=-(1 / 2) g^{\alpha \beta}  \nabla_{\alpha} \phi \nabla_{\beta} \phi$ the kinetic term.

The first term corresponds to the $k$-essence action
\begin{equation}
\begin{split}
\delta S_{\phi} & = \int \mathrm{d}^4 x \sqrt{-g} \frac{\delta P(\phi, X)}{\delta \phi} \delta \phi\,, \\
& = \int \mathrm{d}^4 x \sqrt{-g}\left[\frac{\partial P(\phi, X)}{\partial \phi} \delta \phi \right.
\left.+\frac{\partial P(\phi, X)}{\partial X} \frac{\partial X}{\partial\left(\nabla_{\mu} \phi\right)} \delta\left(\nabla_{\mu} \phi\right)\right]\,.
\end{split}
\end{equation}

Since
\begin{equation}
 \frac{\partial X}{\partial(\nabla_{\mu}\phi)}=-\frac{1}{2}g^{\alpha\beta} \left(\delta^{\mu}_{\alpha}\nabla_{\beta}\phi+\delta^{\mu}_{\beta}\nabla_{\alpha}\phi\right) = - g^{\mu\alpha}\nabla_{\alpha}\phi\,,
\end{equation}
we arrive at
\begin{equation}
\delta S_{\phi} = \int \mathrm{d}^4 x \sqrt{-g}\left[P_{, \phi} \delta \phi - P_{, X} g^{\mu \alpha} \nabla_{\alpha} \phi \delta\left(\nabla_{\mu} \phi\right)\right] \,.
\end{equation}

If we integrate the last term by parts and consider the relation $\delta\left(\nabla_{\mu} \phi\right)=\nabla_{\mu}(\delta \phi)$ we have 
\begin{equation}
\begin{split}
\delta S_{\phi} &=\int \mathrm{d}^4 x \sqrt{-g}\left[P_{, \phi} \delta \phi-\nabla_{\mu}\left(P_{, X} g^{\mu \alpha} \nabla_{\alpha} \phi \delta \phi\right) \right.
\left.+\nabla_{\mu}\left(P_{, X} g^{\mu \alpha} \nabla_{\alpha} \phi\right) \delta \phi\right]\,.
\end{split}
\end{equation}

Let us now assume that the field vanishes at the boundary, that is, at infinity. If we then apply Stokes theorem we have that the term in the middle vanishes. Thus
\begin{equation}\label{boundary}
\begin{split}
\delta S_{\phi}  & = \int d^4 x \sqrt{-g}\left[P_{,\phi}+\nabla_{\mu}\left(P_{,X}g^{\mu \alpha}\nabla_{\alpha}\phi\right)\right] \delta \phi \,,  \\
& =\int \mathrm{d}^4 x \sqrt{-g} \left[P_{,\phi}+\left(P_{,X\phi}\nabla_{\mu}\phi \right.\right.
\left.+P_{,XX}\nabla_{\mu}X\right)g^{\mu \alpha}\nabla_{\alpha}\phi
\left.+P_{,X}g^{\mu \alpha}\nabla_{\mu}\nabla_{\alpha}\phi\right] \delta \phi\,,
\end{split}
\end{equation}
where we have considered the metricity condition $\nabla_{\mu}g^{\mu\alpha}=0$. Further, as the metric is symmetric ($g^{\gamma\beta}=g^{\beta\gamma}$) we can write
\begin{equation}
\begin{split}
\nabla_{\mu} X & = -\frac{1}{2} g^{\gamma \beta}\left(\nabla_{\mu} \nabla_{\gamma} \phi \nabla_{\beta} \phi+\nabla_{\gamma} \phi \nabla_{\mu} \nabla_{\beta} \phi\right)\,,\\
& =-g^{\gamma \beta} \nabla_{\gamma} \phi \nabla_{\mu} \nabla_{\beta} \phi\,.
\end{split}
\end{equation}

If we now use $\square \phi=g^{\mu \alpha} \nabla_{\mu} \nabla_{\alpha} \phi$, where $\square$ is the d'Alembert operator, we have
\begin{equation}\label{alembert}
\begin{split}
\delta S_{\phi} &= \int \mathrm{d}^4 x \sqrt{-g}\left(P_{, \phi}+P_{, X \phi} g^{\mu \alpha} \nabla_{\mu} \phi \nabla_{\alpha} \phi \right. \\
 &\left.-P_{, X X} g^{\gamma \beta} g^{\mu \alpha} \nabla_{\alpha} \phi \nabla_{\gamma} \phi \nabla_{\mu} \nabla_{\beta} \phi+P_{, X} \square \phi\right) \delta \phi\,, \\
&= \int \mathrm{d}^4 x \sqrt{-g}\left(P_{, \phi}+P_{, X} \square \phi-2 P_{, X \phi} X \right.\\
&\left.-P_{, X X} \partial^{\alpha} \phi \partial_{\beta} \phi_{\alpha} \partial^{\beta} \phi\right) \delta \phi\,.
\end{split}
\end{equation}

Let us now turn our attention to the second term in Eq.~\eqref{action}
\begin{equation}
\begin{split}
\delta S_{\rm m} &= \int \mathrm{d}^4 x \sqrt{-g} \frac{\delta\left[f(\phi, X) \tilde{\mathcal{L}}_{\rm m}\right]}{\delta \phi} \delta \phi \,,\\
&= \int \mathrm{d}^4 x \sqrt{-g}\left[f_{, \phi} \delta \phi \tilde{\mathcal{L}}_{\rm m}+f_{, X} \frac{\partial X}{\partial\left(\nabla_{\mu} \phi\right)} \delta\left(\nabla_{\mu} \phi\right) \tilde{\mathcal{L}}_{\rm m}\right] \,,\\
&= \int \mathrm{d}^4 x \sqrt{-g}\left[f_{, \phi} \delta \phi \tilde{\mathcal{L}}_{\rm m}-f_{, X} g^{\mu \alpha} \nabla_{\alpha} \phi \delta\left(\nabla_{\mu} \phi\right) \tilde{\mathcal{L}}_{\rm m}\right]\,.
\end{split}
\end{equation}

If we integrate the last term by parts, this yields
\begin{equation}
\begin{split}
\delta S_{\rm m}&= \int \mathrm{d}^4 x \sqrt{-g}\left[f_{, \phi} \delta \phi \tilde{\mathcal{L}}_{\rm m}-\nabla_{\mu}\left(f_{, X} g^{\mu \alpha} \nabla_{\alpha} \phi \delta \phi \tilde{\mathcal{L}}_{\rm m}\right) \right.\\
&\left.+\nabla_{\mu}\left(f_{, X} g^{\mu \alpha} \nabla_{\alpha} \phi \tilde{\mathcal{L}}_{\rm m}\right) \delta \phi\right]\,,
\end{split}
\end{equation}
whose middle terms vanish if we make use of Stokes theorem, 
\begin{equation}
\begin{split}
\delta S_{\rm m} &= \int \mathrm{d}^4 x \sqrt{-g}\left\{\tilde{\mathcal{L}}_{\rm m}\left[f_{, \phi}+\nabla_{\mu}\left(f_{, X} g^{\mu \alpha} \nabla_{\alpha} \phi\right)\right] \delta \phi \right.\\
&\left.+f_{, X} g^{\mu \alpha} \nabla_{\alpha} \phi \nabla_{\mu} \tilde{\mathcal{L}}_{\rm m} \delta \phi\right\}\,.
\end{split}
\end{equation}

We can see here that, with $P \rightarrow f$, the first term inside the square brackets is identical to the one in Eq.~\eqref{boundary}. Therefore, we can, in a similar fashion, substitute it by considering Eq.~\eqref{alembert}
\begin{equation}\label{substitution}
\begin{split}
\delta S_{\rm m} & = \int \mathrm{d}^4 x \sqrt{-g}\left\{\tilde{\mathcal{L}}_{\rm m}\left[f_{, \phi}+f_{, X} \square \phi-2 f_{, X \phi} X \right.\right.\\
&\left.\left.-f_{, X X} \partial^{\alpha} \phi \partial_{\beta} \phi \nabla_{\alpha} \partial^{\beta} \phi\right]+f_{, X} \partial^{\mu} \phi \nabla_{\mu} \tilde{\mathcal{L}}_{\rm m}\right\} \delta \phi\,.
\end{split}
\end{equation}

If we collect the results for both the variation of the scalar and the matter Lagrangians, Eqs.~(\ref{alembert}) and (\ref{substitution}), respectively, and we further assume that the action is stationary for every variation $\delta \phi$, we shall arrive at the equation of motion for $\phi$
\begin{equation}\label{eom}
\begin{split}
& P_{, \phi}+P_{, X} \square \phi-2 P_{, X \phi} X-P_{, X X} \partial^{\alpha} \phi \partial_{\beta} \phi \nabla_{\alpha} \partial^{\beta} \phi  \\
&=-\tilde{\mathcal{L}}_{\rm m}\left(f_{, \phi}+f_{, X} \square \phi-2 f_{, X \phi} X-f_{, X X} \partial^{\alpha} \phi \partial_{\beta} \phi \nabla_{\alpha} \partial^{\beta} \phi\right) \\
&-f_{, X} \partial^{\mu} \phi \nabla_{\mu} \tilde{\mathcal{L}}_{\rm m}\,.
\end{split}
\end{equation}

If we define $\mathcal{L}_{\rm m}=f \tilde{\mathcal{L}}_{\rm m}$ and substitute the term
\begin{align}
\nabla_{\mu} \tilde{\mathcal{L}}_{\rm m} = &\, \nabla_{\mu}\left(\frac{\mathcal{L}_{\rm m}}{f}\right) = \\
= &\, \frac{\nabla_{\mu} \mathcal{L}_{\rm m}}{f}+\frac{\mathcal{L}_{\rm m}}{f}\left(\frac{f_{, X}}{f} \partial_{\alpha} \phi \nabla_{\mu} \partial^{\alpha} \phi-\frac{f_{, \phi}}{f} \partial_{\mu} \phi\right)\,, \nonumber
\end{align}
in the RHS of Eq.~(\ref{eom}), we may write
\begin{equation}\label{finalA}
P_{, \phi}+P_{, X} \square \phi-2 P_{, X \phi} X-P_{, X X} A = \mathcal{L}_{\rm m} Q\,,    
\end{equation}
being 
\begin{align}
Q = &-\frac{f_{, \phi}}{f}-\frac{f_{, X}}{f}\left(\square \phi+\partial^{\alpha} \phi \frac{\nabla_{\alpha} \mathcal{L}_{\rm m}}{\mathcal{L}_{\rm m}}+\frac{f_{, X}}{f} A+2 \frac{f_{, \phi}}{f} X\right) \nonumber \\
&+2 \frac{f_{, X \phi}}{f} X+\frac{f_{, X X}}{f} A   
\end{align}
and $A=\partial^{\alpha} \phi \partial_{\beta} \phi \left(\nabla_{\alpha} \partial^{\beta}\phi\right)$.

\section{FLRW background}

Let us consider an homogeneous tachyon scalar field $\phi=\phi(t)$ and a FLRW metric. We may then have, at the background level
\begin{subequations}
 \begin{align}
  X = &\, -\frac{1}{2} g^{\alpha \beta} \partial_{\alpha} \phi \partial_{\beta} \phi = \frac{1}{2} \dot{\phi}^{2}\,, \\
  \square \phi = &\, g^{\alpha \beta} \nabla_{\alpha} \nabla_{\beta} \phi = -\ddot{\phi}-3 H \dot{\phi}\,, \\
 A = &\, \phi^{2} \ddot{\phi}\,.
 \end{align}
\end{subequations}

Hence, we may write Eq.~(\ref{finalA}), the equation of motion, as 
\begin{equation}
P_{, \phi}-P_{, X}(\ddot{\phi}+3 H \dot{\phi})-P_{, X \phi} \dot{\phi}^{2}-P_{, X X} \dot{\phi}^{2} \ddot{\phi}=\mathcal{L}_{\rm m} Q\,,
\end{equation}
where 
\begin{equation}
\begin{split}
Q&=-\frac{f_{, \phi}}{f}-\frac{f_{, X}}{f}\left(-\ddot{\phi}-3 H \phi+\dot{\phi}_{m} \frac{\mathcal{L}_{m}}{\mathcal{L}_{m}}+\frac{f_{, X}}{f} \dot{\phi}^{2} \ddot{\phi} \right.\\
&\left.+\frac{f_{, \phi}}{f} \dot{\phi}^{2}\right) +\frac{f_{, X \phi}}{f} \dot{\phi}^{2}+\frac{f_{, X X}}{f} \dot{\phi}^{2} \ddot{\phi}\,.
\end{split}    
\end{equation}

\bibliography{main}

\end{document}